\documentstyle[11pt]{article}

\newcommand{\be}{\begin{equation}}
\newcommand{\ee}{\end{equation}}
\newcommand{\bc}{\begin{center}}
\newcommand{\ec}{\end{center}}
\newcommand{\ba}{\begin{array}}
\newcommand{\ea}{\end{array}}

\small
\begin{document}
\begin{titlepage}

\title{
{\bf The giant resonances in hot nuclei: linear response calculations}
}
\author{ {\bf F\'abio L. Braghin}\thanks{Doctoral fellow of Coordena\c c\~ao de
 Aperfei\c coamento de Pessoal de N\'\i vel Superior,Brasil },
{\bf Dominique  Vautherin}, \\
{\normalsize
Division de Physique Th\'eorique\thanks{Unit\'e de Recherche des
Universit\'es Paris XI et Paris  VI associ\'ee au CNRS} ,
 Institut de Physique Nucl\'eaire,}\\
{\normalsize 91406 Orsay Cedex, France }
\\
{\bf Abdellatif  Abada} \\
{\normalsize   Institute for Theoretical Physics, T\"ubingen University, }\\
{\normalsize   Auf der Morgenstelle  14, D-72076  T\"ubingen, Germany}
}

\date{}
\maketitle
\begin{abstract}
  We calculate the isovector response function of hot nuclear matter
using various effective Skyrme interactions.
For Skyrme forces with a small effective mass
the strength distribution is found to be nearly independent of temperature, and
shows little collective effects.
In contrast effective forces with an effective mass close to
unity produce sizeable collective effects at zero temperature
which disappear at temperatures of a few MeV. We discuss the relevance of
these results for the saturation of the multiplicity of photons
emitted by the giant dipole resonance in hot nuclei beyond $T$=3 MeV observed
in recent experiments.
\end{abstract}
\vskip 1cm

\noindent IPNO/TH 94-95 \hfill{January 1995}

\noindent UNITU-THEP-24/1994

\vskip 1.5cm
\noindent {\it Invited Contribution to the Third IN2P3-Riken Symposium on
Heavy Ion Collisions, Shinrin-Koen, Saitama, Japan, October 24-28, 1994.}
\end{titlepage}

\newpage
\setcounter{equation}{0}

The purpose of the present paper is to investigate the interaction dependence
of the response function of hot nuclear matter to a small isovector
external interaction.
The motivation for such a study is to clarify why some
calculations of these resonances using Random Phase Approximation (RPA)
predict a sharp peak at zero temperature and an increase of
the width with temperature \cite{VAUTHERIN,NVINHMAU}
whereas such is not the case for other self consistent RPA calculations
\cite{SAGAWA}.
In this last case one already has at zero temperature a broad resonance
which shows
little evolution with temperature.  In reference \cite{BRAGHIN}
it was found that the collectivity and temperature dependance of the nuclear
matter response is quite sensitive to the strength of the particle-hole
interaction. This result was however obtained in the case of a schematic Skyrme
type interaction which contains no momentum dependent terms and therefore
cannot be considered as a reliable effective force. For this reason it appears
necessary to perform again such a calculation using a full Skyrme force.
Although calculations for a complete force are somewhat cumbersome we will
show that analytical and transparent formulae can still be obtained. These
formulae generalize those of Garcia- Reccio {\it et al.}
\cite{GARCIA} for the zero temperature response function.

To calculate the response function at finite temperature in nuclear matter, we
consider an infinitesimal external field of the form

\be \label{1}
V_{\hbox{ext}}=
 \epsilon \tau_3 e^{-i {\bf q}.{\bf r} } e^{-i (\omega +i \eta) t}~~,
\ee
where $\tau_3$ is the third isospin Pauli matrix and $\eta$ is a vanishingly
small positive number corresponding to an
adiabatic switching on of the field from the time $t=-\infty$.
The external field
(\ref{1}) induces a difference $\delta \rho$ between the neutron and proton
density distributions which exhibits the same space time dependence
\be \label{2}
<{\bf r}| \delta \rho|{\bf r}> =
\alpha  e^{-i {\bf q}.{\bf r} } e^{-i (\omega +i \eta) t}.
\ee
The response function (or strictly speaking the retarded response function) is
determined by the corresponding polarizability, i.e., the ratio of the density
change to the field strength (which depends only on $\omega$ and ${\bf q}$ ):
\be \label{3}
\Pi(\omega, {\bf q})=\alpha / \epsilon
\ee
In the case of the Skyrme effective force the retarded response function is
found to be of the following form :
\be \label{4}
\Pi(\omega, {\bf q}) = \frac{\Pi_{0} (\omega,{\bf q}) }
{1- V_0 \Pi_{0}(\omega, {\bf q})
-V_1 I_1 (\omega, {\bf q})
-V_2 I_2 (\omega, {\bf q})  }.
\ee
In this formula $V_0, V_1, V_2$ are related to the parameters $t_0, t_1, t_2,
t_3, x_0$ of the Skyrme interaction \cite{VAUBRI,BEFLO,SCHUCK} via the
following formulae
\be \label{5} \ba{ll}
&{\displaystyle
V_0=- \frac{t_0}{2} (x_0+ \frac{1}{2}) - \frac{t_3}{8} \rho_0+
\frac{5}{48} (t_2 - t_1) \frac{\tau_0}{\rho_0} }~,\\
&{\displaystyle V_1= \frac{1}{8} (t_2-  t_1)}~, \\
&{\displaystyle V_2= \frac{1}{16} (t_2- 5 t_1)~, }
\ea
\ee
where $\rho_0$ is the equilibrium density of nuclear matter and where $\tau_0$
is the corresponding value of the kinetic energy density
\be \label{6}
\tau({\bf r})= \sum_i |\nabla \varphi_i({\bf r})|^2.
\ee
In Eq. (\ref{4}) $\Pi_0$ is the unperturbed response function (often
referred to as the Lindhard function \cite{WALECKA}):
\be \label{7}
\Pi_0(\omega, {\bf q})= \frac{2}{ (2 \pi)^3}
\int \hbox {d}^3 k \frac
{ f({\bf k} + {\bf q})- f( {\bf k})  }
{\omega +i \eta-  \epsilon ({\bf k}) + \epsilon({\bf k}+ {\bf q})  },
\ee
where $\epsilon({\bf k})$ is the single particle energy and $f({\bf k}) =
1/(1+\exp{\beta(\epsilon({\bf k})-\mu)})$ the associated occupation number,
$\beta = 1/T$ being the inverse temperature and $\mu$ the chemical potential.
We have used the standard units $\hbar=c=1$.
The functions $I_1$ and $I_2$ appearing in Eq. (\ref{4})
are the generalized Lindhard functions defined by
\be \label{8}
I_1 (\omega, {\bf q})= \frac{1}{ (2 \pi)^3}
\int \hbox {d}^3k ~{\bf k.q}~ \frac
{ f({\bf k} + {\bf q})- f( {\bf k})  }
{\omega +i \eta-  \epsilon ({\bf k}) + \epsilon({\bf k}+ {\bf q})  }~,
\ee
\be \label{9}
I_2 (\omega, {\bf q})= \frac{1}{ (2 \pi)^3}
\int \hbox {d}^3k ~ k^2~ \frac
{ f({\bf k} + {\bf q})- f( {\bf k})  }
{\omega +i \eta-  \epsilon ({\bf k}) + \epsilon({\bf k}+ {\bf q})  },
\ee
Analytical expressions for the real and imaginary parts of
$\Pi_0(\omega,{\bf q})$ can be found in reference \cite{BRAGHIN}.
For instance the imaginary part is given by
\be \label{9A}
{\displaystyle
 \Im m ~\Pi_{0}(\omega, {\bf q}) = -\frac{m^{*2}}{2 \pi q\beta}
 \log \frac{1+ e^{\beta (\mu-E_-)}}{1+e^{\beta(\mu-E_+)}} }
\ee
where $m^*$ is the effective mass in nuclear matter calculated with the Skyrme
force \cite{SCHUCK}, while
\be \label{21}
E_{\pm}=\frac{m^*}{2 q^2} (\omega \pm \frac{q^2}{2 m^*})^2.
\ee
The real part can be expressed as:
\be \label{22}
{\displaystyle \Re e ~\Pi_{0} = - \int
\frac{m^*k}{2 \pi^{2}}  \left\{ -1 +
\frac{k}{2q}\left( \phi( \frac{m^*\omega}{kq}+ \frac{q}{2k})-
\phi( \frac{m^*\omega}{ kq}-\frac{q}{2k})\right) \right\} ~~\hbox {d} f(k,T) }
\ee
with:
\be \label{24}
\phi (x)= (1-x^2) \log \left | \frac{x-1}{x+1} \right | ~.
\ee
 From the previous equations the value of $I_1$ can subsequently
be calculated from the identity
\be \label{10}
m^* ( \omega + \epsilon({\bf q}) ) \Pi_0 + 2 I_1(\omega,{\bf q})=0 ~.
\ee
In the case of zero temperature, expressions for $I_2$ can be
found in Ref. \cite{GARCIA}. For non zero temperature the calculation of
the integral (\ref{9}) is similar to that of  $\Pi_0$. The result is that
the real
part of $I_2$ is an average of the zero temperature values calculated for
the same energy $\omega$ and momentum $q$, but with various
Fermi momenta $k_F$ distributed with a weight factor which is just the
derivative of the Fermi occupation number. Explicitely on has the following
formula:
\be \label{11}
\Re e ~I_2(\omega,q,T)= -\int \Re e~I_2(\omega,q,T=0,k_F=k)~\hbox {d}f(k,T).
\ee
The imaginary part of $I_2$ is given by
\be \label{12}
 \Im m I_2(\omega , q) = -\frac{m^{*3}}{2 \pi q \beta^2} X,
\ee
with
\be \label{11a}
X=\beta E_+ \log \frac{1+ e^{\beta (\mu -E_-)}}
{1+e^{\beta(\mu - E_+)}}
+Li_2(1+ e^{\beta (\mu -E_+)}) - Li_2(1+ e^{\beta (\mu -E_-)}) ,
\ee
where $Li_2$ the Euler dilogarithmic function \cite{ABRAMOWITZ} and
$E_{\pm}$ has been defined in (\ref{21}).

The imaginary part of $\Pi$ (c.f., Eq. (\ref{4})) is related to
the value of the strength per unit volume $S(\omega)$ for the operator $\tau_3
\exp(i {\bf q}.{\bf r})$  by
\be \label{14}
S(\omega)= -\frac{1}{\pi} \Im m \Pi(\omega,q).
\ee
It satisfies the following sum rule
\be \label{15}
\int^{\infty}_{0} S(\omega) \omega d\omega = \frac{q^2 }{2 m^*} \rho_0 ( 1 +
\kappa )
\ee
\noindent where $ \kappa $ is the enhancement factor due to the momentum
dependent terms of the Skyrme interaction:
\be
\kappa = \frac{ 2 m k_F^3}{ 3  \pi^2}( t_1 + t_2 )
\ee
The strength distribution as a function of the frequency $\omega$
is plotted for the three Skyrme forces SV, SI and  SII in figures 1, 2  and 3
respectively, for various values of the temperature.
In these figures the value of the momentum transfer was chosen
to correspond to the dipole mode in lead-208 in the Steinwedel- Jenssen model.
In this model neutrons oscillate against protons inside a sphere of radius $R$
(the nuclear radius) according to the formula \cite{SCHUCK}
\be \label{16}
<{\bf r}| \delta \rho|{\bf r}>= \varepsilon
\sin ({\bf q}.{\bf r}) \sin( \omega t),
\ee
where
\be \label{17}
q = \frac{\pi}{2 R}.
\ee
 The radius of lead-208 is $R$=6.7 fm. This value leads to  $q$=0.23 fm$^{-1}$.

It can be noted in Fig.1 that in the case of the Skyrme force SV, the dipole
strength is nearly independent of temperature.
We found this to be the case for all Skyrme forces having small value of the
effective mass, such SV (m$^*$=0.36, in units
of the bare nucleon mass whenever a numerical value is given)  and
SIV (m$^*$=0.47) \cite{BEFLO}.
These forces also give an energy of the dipole resonance in
lead-208 much higher (about 25- 30 MeV) than the observed value (15 MeV). In
contrast effective forces with an effective mass of the order of unity such as
SI (m$^*$=0.91) and SVI (m$^*$=0.95) \cite{VAUBRI}, give a sharper resonance at
zero temperature, a good agreement with the observed position and also
a stronger temperature dependence. This result  may be related to
the saturation of the multiplicity of photons
emitted by the giant dipole resonance in hot nuclei when their temperature
exceeds $T$=3 MeV observed in some experiments \cite{LEFAOU}.

Concerning the collectivity of the resonance one can see from Fig.~2 that
the SI Skyrme force gives a more concentrated coherent behavior.
Taking forces with
smaller value of effective mass, for example SII (m$^*$ = 0.58)  in Fig.~3,
the interaction ceases to be important and the response function has
pratically the form of the bare response, i.e., the imaginary part of
the unperturbed Lindhard function. With a smaller effective mass ( as
in SIV and SV), the collectivity is distributed over an even larger range of
frequencies.

It is worthwhile emphasizing that all Skyrme interactions considered here
provide a good
description of nuclear ground state properties such as root mean square radii
and binding energies. Obviously such a good description is not sufficient to
garantee at the same time an accurate description of the isovector dipole
response function:  even with a good fit to the symmetry energy coefficient
there is a large dispersion in the results for the nuclear matter isovector
response function.

Our main conclusion is thus that, even at the level of the random phase
approximation,  uncertainties in the predictions of microscopic
calculations do remain for the dipole strength and its temperature dependence.
We find that the effective mass mainly determines the energy of the
resonance, its collectivity and its evolution
with temperature. However other effects are known to be important in the
description of the observed width such as the coupling to two particle-
two hole configurations \cite{ADACHI}.
A larger number of experimental data on the temperature dependence of the
strength would be highly desirable to resolve the remaining uncertainties.
Simultaneously further microscopic theoretical studies of the effects not
included in  RPA would be necessary.

\vskip 0.3cm
\noindent {\Large {\bf Acknowledgements}}
\vskip 0.2cm

We are grateful to Nguyen Van Giai and Nicole Vinh Mau
for stimulating discussions.


\vskip 0.1cm
\begin{thebibliography}{11}
\bibitem{VAUTHERIN} D. Vautherin and N. Vinh Mau, Nucl. Phys. {\bf A422}
(1984) 140
\bibitem{NVINHMAU} N. Vinh Mau, Nucl. Phys. {\bf A548} (1992) 381
\bibitem{SAGAWA} H. Sagawa and G. F. Bertsch, Phys. Lett. {\bf B 146} (1984)
140
\bibitem{BRAGHIN} F. L. Braghin and D. Vautherin, Phys. Lett. {\bf B 333}
(1994) 289
\bibitem{GARCIA} C. Garcia- Recio, J. Navarro, N. Van Giai and
L. L. Salcedo, Ann. of Phys. {\bf 214} (1992) 293
\bibitem{VAUBRI} D. Vautherin and D. M. Brink, Phys. Rev.  {\bf C 5} (1972)
 626
\bibitem{BEFLO} M. Beiner, H. Flocard, N. Van Giai and P. Quentin, Nuc. Phys.
{\bf A238} (1975) 29
\bibitem{SCHUCK} P. Ring and P. Schuck, {\it The nuclear many body problem},
Springer, Berlin, 1982
\bibitem{WALECKA} A. L. Fetter and J. D. Walecka, {\it Quantum Theory of
many particle systems}, Mc Graw Hill, New York, 1971
\bibitem{ABRAMOWITZ} M. Abramowitz and I. Stegun, Handbook of Mathematical
Functions, Dover, New- York, 1965
\bibitem{LEFAOU} H. Lefaou {\it et al}, Phys. Rev. Lett. {\bf 72} (1994) 289
\bibitem{ADACHI} S. Adachi and N. Van Giai, Phys. Lett.  {\bf 149B}
(1984) 447
\end{thebibliography}
\end{document}